\documentclass[zpreprint,zbstpl]{zeus_paper} % for PLB

\usepackage[english]{babel}

\newcommand{\Zdetdesc}{%
A detailed description of the ZEUS detector can be found 
elsewhere~\cite{zeus:1993:bluebook}. A brief outline of the 
components that are most relevant for this analysis is given
below.\xspace}
\newcommand{\Zsttdesc}[1]{%
The STT consisted of 48 sectors of two different sizes. Each sector
contained 192 (small sector) or 264 (large sector) straws of diameter
7.5 mm arranged into 3 layers. The sectors were trapezoidal in shape
and each subtended an azimuthal angle of $60^{\circ}$ -- 6 sectors
formed a so-called superlayer. A particle passing through the complete
detector traversed 8 superlayers, which were rotated around the beam
direction at angles of $30^{\circ}$ or $15^{\circ}$ to each other. The STT
covered the polar-angle region $5^{\circ}<\theta<23^{\circ}$.
}
\newcommand{\Zcaldesc}{%
The high-resolution uranium--scintillator calorimeter (CAL)~\citeCAL consisted 
of three parts: the forward (FCAL), the barrel (BCAL) and the rear (RCAL)
calorimeters. Each part was subdivided transversely into towers and
longitudinally into one electromagnetic section (EMC) and either one (in RCAL)
or two (in BCAL and FCAL) hadronic sections (HAC). The smallest subdivision of
the calorimeter is called a cell.  The CAL energy resolutions, as measured under
test-beam conditions, were $\sigma(E)/E=0.18/\sqrt{E}$ for electrons and
$\sigma(E)/E=0.35/\sqrt{E}$ for hadrons ($E$ in $\Gev$).}
\chardef\usc=95
\chardef\til=126
\catcode`\@=11 % @ signs are now treated as letters
\DeclareRobustCommand\xdotspace{\futurelet\@let@token\@xdotspace}
\def\@xdotspace{%
  \ifx\@let@token.\else
  \ifx\@let@token\bgroup.\else
  \ifx\@let@token\egroup.\else
  \ifx\@let@token\/.\else
  \ifx\@let@token\ .\else
  \ifx\@let@token~.\else
  \ifx\@let@token!.\else
  \ifx\@let@token,.\else
  \ifx\@let@token:.\else
  \ifx\@let@token;.\else
  \ifx\@let@token?.\else
  \ifx\@let@token/.\else
  \ifx\@let@token'.\else
  \ifx\@let@token).\else
  \ifx\@let@token-.\else
  \ifx\@let@token\@xobeysp.\else
  \ifx\@let@token\space.\else
  \ifx\@let@token\@sptoken.\else
   .\space
   \fi\fi\fi\fi\fi\fi\fi\fi\fi\fi\fi\fi\fi\fi\fi\fi\fi\fi}
\catcode`\@=12 % @ signs are no longer letters

\newcommand{\stru}[2]{%
   \relax\ifmmode\hbox{\vrule height#1 depth#2 width0pt}%
   \else\vrule height#1 depth#2 width0pt\fi}
\newcommand{\Ronum}[1]{\uppercase\expandafter{\romannumeral#1}}
\newcommand{\ronum}[1]{\expandafter{\romannumeral#1}}
\DeclareRobustCommand{\LaTeXZ}{%
  \LaTeX\kern-.05em4\kern-.1em
  {\raisebox{-0.2ex}{$\scriptstyle\text{ZEUS}$}}\xspace}
\newcommand{\eq}[1]{(\ref{eq-#1})}
\newcommand{\eqsand}[2]{(\ref{eq-#1}) and~(\ref{eq-#2})}

\newcommand{\fig}[1]{Fig.~\ref{fig-#1}}
\newcommand{\Fig}[1]{Figure~\ref{fig-#1}}

\newcommand{\tab}[1]{Table~\ref{tab-#1}}

\newcommand{\Sect}[1]{Section~\ref{sec-#1}}

\DeclareMathAlphabet{\mathbf}{OT1}{cmr}{bx}{sl}
\newcommand{\eVdist}{\kern-0.06667em}

\newcommand{\Gev}{{\text{Ge}\eVdist\text{V\/}}}

\newcommand{\mev}{{\,\text{Me}\eVdist\text{V\/}}}
\newcommand{\gev}{{\,\text{Ge}\eVdist\text{V\/}}}

\newcommand{\nbi}{\,\text{nb}^{-1}}

\newcommand{\cm}{\,\text{cm}}

\newcommand{\ns}{\,\text{ns}}

\newcommand{\Tesla}{\,\text{T}}

\newcommand{\slashfrac}[2]{%
  \raisebox{0.5ex}{\ensuremath #1}\kern-0.12em/\kern-0.08em
  \raisebox{-.8ex}{\ensuremath #2}}
\newcommand{\sqr}[3]{%
    {\vcenter{\hrule height.#3ex\hbox{\vrule width.#2ex height#1ex
     \kern#1ex\vrule width.#3ex}\hrule height.#2ex}}}

\catcode`\@=11 % @ signs are now treated as letters
\newcommand{\parenbar}{\mathpalette\p@renb@r}
\def\p@renb@r#1#2{\vbox{%
  \ifx#1\scriptscriptstyle \dimen@.7em\dimen@ii.2em\else
  \ifx#1\scriptstyle \dimen@.8em\dimen@ii.25em\else
  \dimen@1em\dimen@ii.4em\fi\fi \offinterlineskip
  \ialign{\hfill##\hfill\cr
    \vbox{\hrule width\dimen@ii}\cr
    \noalign{\vskip-.3ex}%
    \hbox to\dimen@{$\mathchar300\hfil\mathchar301$}\cr
    \noalign{\vskip-.3ex}%
    $#1#2$\cr}}}
\catcode`\@=12 % @ signs are no longer letters

\newcommand{\IP}{{\rm I$\kern-0.01667em$P}\xspace}

\newcommand{\Lumi}{{\cal L}}

\mathchardef\qsm=63
\mathchardef\pls=43
\mathchardef\mns=512
\mathchardef\plm=518
\mathchardef\eql=61
\mathchardef\smallleft=300
\mathchardef\smallright=301
\mathchardef\les=316
\mathchardef\gre=318
\mathchardef\leq=532
\mathchardef\grq=533
\catcode`\@=11 % @ signs are now treated as letters
\newcounter{pict@width}
\newcounter{pict@height}
\newlength{\pict@scale}
\newcommand{\psfigadd}[4]{%
\setcounter{pict@width}{1*\ratio{#2+\pict@scale/2}{\pict@scale}}
\setcounter{pict@height}{1*\ratio{#3+\pict@scale/2}{\pict@scale}}
\setlength{\unitlength}{\pict@scale}
\hbox to #2{\hspace{-\fill}\begin{picture}(\thepict@width,\thepict@height)
\put(0,0){\psfig{figure=#1,width=#2,height=#3,clip=}}
\SetScale{0.283466457}
\SetWidth{1.763889}
{#4}
\end{picture}}
}
\newcounter{pict@widthfst}
\newcounter{pict@widthscd}
\newcounter{pict@widthtot}
\newcommand{\psfigaddtwo}[7]{%
\setcounter{pict@widthfst}{1*\ratio{#2+\pict@scale/2}{\pict@scale}}
\setcounter{pict@widthscd}{1*\ratio{#2+#4+\pict@scale/2}{\pict@scale}}
\setcounter{pict@widthtot}{1*\ratio{#2+#4+#6+\pict@scale/2}{\pict@scale}}
\setcounter{pict@height}{1*\ratio{#3+\pict@scale/2}{\pict@scale}}
\setlength{\unitlength}{\pict@scale}
\hbox{\hspace{-\fill}\begin{picture}(\thepict@widthtot,\thepict@height)
\put(0,0){\psfig{figure=#1,width=#2,height=#3,clip=}}
\put(\thepict@widthscd,0){\psfig{figure=#5,width=#6,height=#3,clip=}}
\SetScale{0.283466457}
\SetWidth{1.763889}
{#7}
\end{picture}}
}
\newcommand{\psfigror}[4]{%
\setcounter{pict@width}{1*\ratio{#2+\pict@scale/2}{\pict@scale}}
\setcounter{pict@height}{1*\ratio{#3+\pict@scale/2}{\pict@scale}}
\setlength{\unitlength}{\pict@scale}
\hbox{\begin{picture}(\thepict@width,\thepict@height)
\put(0,\thepict@height){\psfig{figure=#1,width=#3,height=#2,clip=,angle=270}}
\SetScale{0.283466457}
\SetWidth{1.763889}
{#4}
\end{picture}}
}
\newcommand{\psfigrol}[4]{%
\setcounter{pict@width}{1*\ratio{#2+\pict@scale/2}{\pict@scale}}
\setcounter{pict@height}{1*\ratio{#3+\pict@scale/2}{\pict@scale}}
\setlength{\unitlength}{\pict@scale}
\hbox{\begin{picture}(\thepict@width,\thepict@height)
\put(0,0){\psfig{figure=#1,width=#3,height=#2,clip=,angle=90}}
\SetScale{0.283466457}
\SetWidth{1.763889}
{#4}
\end{picture}}
}
\catcode`\@=12 % @ signs are no longer letters
\newlength\listtextwidth

\catcode`\@=11 % @ signs are now treated as letters
\newlength{\@tabfninsert}
\newlength{\@tabfnwidth}
\newcommand{\tabfootnote}[2]{%
  \setlength{\@tabfninsert}{0.8em}
  \setlength{\@tabfnwidth}{\textwidth}
  \addtolength{\@tabfnwidth}{-\@tabfninsert}
  \addtolength{\@tabfnwidth}{-0.4em}
  \noindent\makebox[\@tabfninsert][r]{\footnotesize$^{#1}$\hfil}\hfill%
  \parbox[t]{\@tabfnwidth}{\footnotesize #2\hfill}}
\catcode`\@=12 % @ signs are no longer letters

\newcommand{\sigtot}{\sigma_{\rm tot}^{\gamma p}}
\newcommand{\st}{$\sigtot$ }

\newcommand {\pom} {I\!\!P}

\newcommand {\reg} {I\!\!R}

\newcommand{\ts}{{\rm TAG6}}
\newcommand{\tss}{{\rm TAG6 }}

\newcommand{\BH}{bremsstrahlung }
\newcommand{\Fts}{F_{\gamma}^{\,\ts}}
\newcommand{\Ftest}{F_{\gamma}^{\,\rm test}}
\newcommand{\Ainc}{A_{\rm inc}}

\newcommand{\myeq}[1]{Eq.~\eq{#1}}
\newcommand{\myeqsand}[2]{Eqs.~\eqsand{#1}{#2}}
\newcommand{\myZcoosysA}{%
The ZEUS coordinate system is a right-handed Cartesian system, with the $Z$
axis pointing in the proton beam direction, referred to as the ``forward
direction'', and the $X$ axis pointing
toward the center of HERA.
The coordinate origin is at the nominal interaction point.\xspace}
\newcommand{\myZcoosysfnA}{\footnote{\myZcoosysA}}
\newcommand{\myZtrackingdesc}[1]{%
Charged particles were tracked
in the central tracking detector (CTD)~\citeCTD and the microvertex
detector (MVD)~\citeMVD. The CTD and the MVD
operated in a magnetic field of $1.43\Tesla$ provided by a thin
superconducting solenoid. The CTD drift chamber covered the
polar-angle#1 region \mbox{$15^\circ<\theta<164^\circ$}. The MVD
silicon tracker provided polar angle coverage for tracks
from $7^\circ$ to $150^\circ$.
}

\def\citeCTD{{\cite{%
nim:a279:290,*npps:b32:181,*nim:a338:254%
}}\xspace}
\def\citeMVD{{\cite{%
nim:a581:656%
}}\xspace}
\def\citeCAL{{\cite{%
nim:a309:77,*nim:a309:101,*nim:a321:356,*nim:a336:23%
}}\xspace}

\begin{document}

\prepnum{{DESY--10--178}}

\title{Measurement of the energy dependence of the total
       photon-proton cross section at HERA}
                    
\author{ZEUS Collaboration}
\date{October 2010}

\abstract{
The energy dependence of the photon-proton total cross section,
\st, was determined from $e^+p$  scattering data collected with the
ZEUS detector at HERA at three values of the center-of-mass energy,
$W$, of the $\gamma p$ system in the range $194 < W < 296\gev$.
This is the first determination of the $W$ dependence of
\st from a single experiment at high $W$. 
Parameterizing \st $\propto W^{2\epsilon}$,
$\epsilon = 0.111 \pm 0.009 \, {\rm (stat.)} \pm 0.036 \, {\rm (syst.)}$
was obtained.
 }

\makezeustitle

\def\3{\ss}

\pagenumbering{Roman}

%------------------------------------------------------------------------------
%       Authors
%------------------------------------------------------------------------------

%===================================================================
%
%  MEMBER NAME  AUTH166 (ZEUS)     M  TEX
%
%  JH.: transformed to a format, which is suited as input for
%       CONVERT, which automatically creates author-indices
%
%  Don't remove lines starting with a percent sign %,
%  CONVERT may need them urgently !
%  
%=====================================================================
                                                   %
\begin{center}
{                      \Large  The ZEUS Collaboration              }
\end{center}

{\small

%  members:

{\mbox H.~Abramowicz$^{44, af}$, }
{\mbox I.~Abt$^{34}$, }
{\mbox L.~Adamczyk$^{13}$, }
{\mbox M.~Adamus$^{53}$, }
{\mbox R.~Aggarwal$^{7, d}$, }
{\mbox S.~Antonelli$^{4}$, }
{\mbox P.~Antonioli$^{3}$, }
{\mbox A.~Antonov$^{32}$, }
{\mbox M.~Arneodo$^{49}$, }
{\mbox V.~Aushev$^{26, aa}$
% ...
}
\newpage

%       institutes:

\makebox[3em]{$^{1}$}
\begin{minipage}[t]{14cm}
{\it Argonne National Laboratory, Argonne, Illinois 60439-4815, USA}~$^{A}$

\end{minipage}\\
\makebox[3em]{$^{2}$}
\begin{minipage}[t]{14cm}
{\it Andrews University, Berrien Springs, Michigan 49104-0380, USA}

\end{minipage}\\
\makebox[3em]{$^{3}$}
\begin{minipage}[t]{14cm}
{\it INFN Bologna, Bologna, Italy}~$^{B}$

\end{minipage}\\
\makebox[3em]{$^{4}$}
\begin{minipage}[t]{14cm}
{\it University and INFN Bologna, Bologna, Italy}~$^{B}$

\end{minipage}\\
\makebox[3em]{$^{5}$}
\begin{minipage}[t]{14cm}
{\it Physikalisches Institut der Universit\"at Bonn,
Bonn, Germany}~$^{C}$

\end{minipage}\\
\makebox[3em]{$^{6}$}
\begin{minipage}[t]{14cm}
{\it H.H.~Wills Physics Laboratory, University of Bristol,
Bristol, United Kingdom}~$^{D}$

\end{minipage}\\
\makebox[3em]{$^{7}$}
\begin{minipage}[t]{14cm}
{\it Panjab University, Department of Physics, Chandigarh, India}

\end{minipage}\\
\makebox[3em]{$^{8}$}
\begin{minipage}[t]{14cm}
{\it Calabria University,
Physics Department and INFN, Cosenza, Italy}~$^{B}$

\end{minipage}\\
\makebox[3em]{$^{9}$}
\begin{minipage}[t]{14cm}
{\it Institute for Universe and Elementary Particles, Chonnam National University,\\
Kwangju, South Korea}

\end{minipage}\\
\makebox[3em]{$^{10}$}
\begin{minipage}[t]{14cm}
{\it Jabatan Fizik, Universiti Malaya, 50603 Kuala Lumpur, Malaysia}~$^{E}$

\end{minipage}\\
\makebox[3em]{$^{11}$}
\begin{minipage}[t]{14cm}
{\it Nevis Laboratories, Columbia University, Irvington on Hudson,
New York 10027, USA}~$^{F}$

\end{minipage}\\
\makebox[3em]{$^{12}$}
\begin{minipage}[t]{14cm}
{\it The Henryk Niewodniczanski Institute of Nuclear Physics, Polish Academy of \\
Sciences, Cracow, Poland}~$^{G}$

\end{minipage}\\
\makebox[3em]{$^{13}$}
\begin{minipage}[t]{14cm}
{\it Faculty of Physics and Applied Computer Science, AGH-University of Science and \\
Technology, Cracow, Poland}~$^{H}$

\end{minipage}\\
\makebox[3em]{$^{14}$}
\begin{minipage}[t]{14cm}
{\it Department of Physics, Jagellonian University, Cracow, Poland}

\end{minipage}\\
\makebox[3em]{$^{15}$}
\begin{minipage}[t]{14cm}
{\it Deutsches Elektronen-Synchrotron DESY, Hamburg, Germany}

\end{minipage}\\
\makebox[3em]{$^{16}$}
\begin{minipage}[t]{14cm}
{\it Deutsches Elektronen-Synchrotron DESY, Zeuthen, Germany}

\end{minipage}\\
\makebox[3em]{$^{17}$}
\begin{minipage}[t]{14cm}
{\it INFN Florence, Florence, Italy}~$^{B}$

\end{minipage}\\
\makebox[3em]{$^{18}$}
\begin{minipage}[t]{14cm}
{\it University and INFN Florence, Florence, Italy}~$^{B}$

\end{minipage}\\
\makebox[3em]{$^{19}$}
\begin{minipage}[t]{14cm}
{\it Fakult\"at f\"ur Physik der Universit\"at Freiburg i.Br.,
Freiburg i.Br., Germany}

\end{minipage}\\
\makebox[3em]{$^{20}$}
\begin{minipage}[t]{14cm}
{\it School of Physics and Astronomy, University of Glasgow,
Glasgow, United Kingdom}~$^{D}$

\end{minipage}\\
\makebox[3em]{$^{21}$}
\begin{minipage}[t]{14cm}
{\it Department of Engineering in Management and Finance, Univ. of
the Aegean, Chios, Greece}

\end{minipage}\\
\makebox[3em]{$^{22}$}
\begin{minipage}[t]{14cm}
{\it Hamburg University, Institute of Experimental Physics, Hamburg,
Germany}~$^{I}$

\end{minipage}\\
\makebox[3em]{$^{23}$}
\begin{minipage}[t]{14cm}
{\it Imperial College London, High Energy Nuclear Physics Group,
London, United Kingdom}~$^{D}$

\end{minipage}\\
\makebox[3em]{$^{24}$}
\begin{minipage}[t]{14cm}
{\it Institute of Particle and Nuclear Studies, KEK,
Tsukuba, Japan}~$^{J}$

\end{minipage}\\
\makebox[3em]{$^{25}$}
\begin{minipage}[t]{14cm}
{\it Institute of Physics and Technology of Ministry of Education and
Science of Kazakhstan, Almaty, Kazakhstan}

\end{minipage}\\
\makebox[3em]{$^{26}$}
\begin{minipage}[t]{14cm}
{\it Institute for Nuclear Research, National Academy of Sciences, and
Kiev National University, Kiev, Ukraine}

\end{minipage}\\
\makebox[3em]{$^{27}$}
\begin{minipage}[t]{14cm}
{\it Kyungpook National University, Center for High Energy Physics, Daegu,
South Korea}~$^{K}$

\end{minipage}\\
\makebox[3em]{$^{28}$}
\begin{minipage}[t]{14cm}
{\it Institut de Physique Nucl\'{e}aire, Universit\'{e} Catholique de Louvain, Louvain-la-Neuve,\\
Belgium}~$^{L}$

\end{minipage}\\
\makebox[3em]{$^{29}$}
\begin{minipage}[t]{14cm}
{\it Departamento de F\'{\i}sica Te\'orica, Universidad Aut\'onoma
de Madrid, Madrid, Spain}~$^{M}$

\end{minipage}\\
\makebox[3em]{$^{30}$}
\begin{minipage}[t]{14cm}
{\it Department of Physics, McGill University,
Montr\'eal, Qu\'ebec, Canada H3A 2T8}~$^{N}$

\end{minipage}\\
\makebox[3em]{$^{31}$}
\begin{minipage}[t]{14cm}
{\it Meiji Gakuin University, Faculty of General Education,
Yokohama, Japan}~$^{J}$

\end{minipage}\\
\makebox[3em]{$^{32}$}
\begin{minipage}[t]{14cm}
{\it Moscow Engineering Physics Institute, Moscow, Russia}~$^{O}$

\end{minipage}\\
\makebox[3em]{$^{33}$}
\begin{minipage}[t]{14cm}
{\it Moscow State University, Institute of Nuclear Physics,
Moscow, Russia}~$^{P}$

\end{minipage}\\
\makebox[3em]{$^{34}$}
\begin{minipage}[t]{14cm}
{\it Max-Planck-Institut f\"ur Physik, M\"unchen, Germany}

\end{minipage}\\
\makebox[3em]{$^{35}$}
\begin{minipage}[t]{14cm}
{\it NIKHEF and University of Amsterdam, Amsterdam, Netherlands}~$^{Q}$

\end{minipage}\\
\makebox[3em]{$^{36}$}
\begin{minipage}[t]{14cm}
{\it Physics Department, Ohio State University,
Columbus, Ohio 43210, USA}~$^{A}$

\end{minipage}\\
\makebox[3em]{$^{37}$}
\begin{minipage}[t]{14cm}
{\it Department of Physics, University of Oxford,
Oxford, United Kingdom}~$^{D}$

\end{minipage}\\
\makebox[3em]{$^{38}$}
\begin{minipage}[t]{14cm}
{\it INFN Padova, Padova, Italy}~$^{B}$

\end{minipage}\\
\makebox[3em]{$^{39}$}
\begin{minipage}[t]{14cm}
{\it Dipartimento di Fisica dell' Universit\`a and INFN,
Padova, Italy}~$^{B}$

\end{minipage}\\
\makebox[3em]{$^{40}$}
\begin{minipage}[t]{14cm}
{\it Department of Physics, Pennsylvania State University, University Park,\\
Pennsylvania 16802, USA}~$^{F}$

\end{minipage}\\
\makebox[3em]{$^{41}$}
\begin{minipage}[t]{14cm}
{\it Polytechnic University, Sagamihara, Japan}~$^{J}$

\end{minipage}\\
\makebox[3em]{$^{42}$}
\begin{minipage}[t]{14cm}
{\it Dipartimento di Fisica, Universit\`a 'La Sapienza' and INFN,
Rome, Italy}~$^{B}$

\end{minipage}\\
\makebox[3em]{$^{43}$}
\begin{minipage}[t]{14cm}
{\it Rutherford Appleton Laboratory, Chilton, Didcot, Oxon,
United Kingdom}~$^{D}$

\end{minipage}\\
\makebox[3em]{$^{44}$}
\begin{minipage}[t]{14cm}
{\it Raymond and Beverly Sackler Faculty of Exact Sciences, School of Physics, \\
Tel Aviv University, Tel Aviv, Israel}~$^{R}$

\end{minipage}\\
\makebox[3em]{$^{45}$}
\begin{minipage}[t]{14cm}
{\it Department of Physics, Tokyo Institute of Technology,
Tokyo, Japan}~$^{J}$

\end{minipage}\\
\makebox[3em]{$^{46}$}
\begin{minipage}[t]{14cm}
{\it Department of Physics, University of Tokyo,
Tokyo, Japan}~$^{J}$

\end{minipage}\\
\makebox[3em]{$^{47}$}
\begin{minipage}[t]{14cm}
{\it Tokyo Metropolitan University, Department of Physics,
Tokyo, Japan}~$^{J}$

\end{minipage}\\
\makebox[3em]{$^{48}$}
\begin{minipage}[t]{14cm}
{\it Universit\`a di Torino and INFN, Torino, Italy}~$^{B}$

\end{minipage}\\
\makebox[3em]{$^{49}$}
\begin{minipage}[t]{14cm}
{\it Universit\`a del Piemonte Orientale, Novara, and INFN, Torino,
Italy}~$^{B}$

\end{minipage}\\
\makebox[3em]{$^{50}$}
\begin{minipage}[t]{14cm}
{\it Department of Physics, University of Toronto, Toronto, Ontario,
Canada M5S 1A7}~$^{N}$

\end{minipage}\\
\makebox[3em]{$^{51}$}
\begin{minipage}[t]{14cm}
{\it Physics and Astronomy Department, University College London,
London, United Kingdom}~$^{D}$

\end{minipage}\\
\makebox[3em]{$^{52}$}
\begin{minipage}[t]{14cm}
{\it Warsaw University, Institute of Experimental Physics,
Warsaw, Poland}

\end{minipage}\\
\makebox[3em]{$^{53}$}
\begin{minipage}[t]{14cm}
{\it Institute for Nuclear Studies, Warsaw, Poland}

\end{minipage}\\
\makebox[3em]{$^{54}$}
\begin{minipage}[t]{14cm}
{\it Department of Particle Physics, Weizmann Institute, Rehovot,
Israel}~$^{S}$

\end{minipage}\\
\makebox[3em]{$^{55}$}
\begin{minipage}[t]{14cm}
{\it Department of Physics, University of Wisconsin, Madison,
Wisconsin 53706, USA}~$^{A}$

\end{minipage}\\
\makebox[3em]{$^{56}$}
\begin{minipage}[t]{14cm}
{\it Department of Physics, York University, Ontario, Canada M3J
1P3}~$^{N}$

\end{minipage}\\
\vspace{30em} \pagebreak[4]

%  references concerning institutes;

\makebox[3ex]{$^{ A}$}
\begin{minipage}[t]{14cm}
 supported by the US Department of Energy\
\end{minipage}\\
\makebox[3ex]{$^{ B}$}
\begin{minipage}[t]{14cm}
 supported by the Italian National Institute for Nuclear Physics (INFN) \
\end{minipage}\\
\makebox[3ex]{$^{ C}$}
\begin{minipage}[t]{14cm}
 supported by the German Federal Ministry for Education and Research (BMBF), under
 contract No. 05 H09PDF\
\end{minipage}\\
\makebox[3ex]{$^{ D}$}
\begin{minipage}[t]{14cm}
 supported by the Science and Technology Facilities Council, UK\
\end{minipage}\\
\makebox[3ex]{$^{ E}$}
\begin{minipage}[t]{14cm}
 supported by an FRGS grant from the Malaysian government\
\end{minipage}\\
\makebox[3ex]{$^{ F}$}
\begin{minipage}[t]{14cm}
 supported by the US National Science Foundation. Any opinion,
 findings and conclusions or recommendations expressed in this material
 are those of the authors and do not necessarily reflect the views of the
 National Science Foundation.\
\end{minipage}\\
\makebox[3ex]{$^{ G}$}
\begin{minipage}[t]{14cm}
 supported by the Polish Ministry of Science and Higher Education as a scientific project No.
 DPN/N188/DESY/2009\
\end{minipage}\\
\makebox[3ex]{$^{ H}$}
\begin{minipage}[t]{14cm}
 supported by the Polish Ministry of Science and Higher Education
 as a scientific project (2009-2010)\
\end{minipage}\\
\makebox[3ex]{$^{ I}$}
\begin{minipage}[t]{14cm}
 supported by the German Federal Ministry for Education and Research (BMBF), under
 contract No. 05h09GUF, and the SFB 676 of the Deutsche Forschungsgemeinschaft (DFG) \
\end{minipage}\\
\makebox[3ex]{$^{ J}$}
\begin{minipage}[t]{14cm}
 supported by the Japanese Ministry of Education, Culture, Sports, Science and Technology
 (MEXT) and its grants for Scientific Research\
\end{minipage}\\
\makebox[3ex]{$^{ K}$}
\begin{minipage}[t]{14cm}
 supported by the Korean Ministry of Education and Korea Science and Engineering
 Foundation\
\end{minipage}\\
\makebox[3ex]{$^{ L}$}
\begin{minipage}[t]{14cm}
 supported by FNRS and its associated funds (IISN and FRIA) and by an Inter-University
 Attraction Poles Programme subsidised by the Belgian Federal Science Policy Office\
\end{minipage}\\
\makebox[3ex]{$^{ M}$}
\begin{minipage}[t]{14cm}
 supported by the Spanish Ministry of Education and Science through funds provided by
 CICYT\
\end{minipage}\\
\makebox[3ex]{$^{ N}$}
\begin{minipage}[t]{14cm}
 supported by the Natural Sciences and Engineering Research Council of Canada (NSERC) \
\end{minipage}\\
\makebox[3ex]{$^{ O}$}
\begin{minipage}[t]{14cm}
 partially supported by the German Federal Ministry for Education and Research (BMBF)\
\end{minipage}\\
\makebox[3ex]{$^{ P}$}
\begin{minipage}[t]{14cm}
 supported by RF Presidential grant N 41-42.2010.2 for the Leading
 Scientific Schools and by the Russian Ministry of Education and Science through its
 grant for Scientific Research on High Energy Physics\
\end{minipage}\\
\makebox[3ex]{$^{ Q}$}
\begin{minipage}[t]{14cm}
 supported by the Netherlands Foundation for Research on Matter (FOM)\
\end{minipage}\\
\makebox[3ex]{$^{ R}$}
\begin{minipage}[t]{14cm}
 supported by the Israel Science Foundation\
\end{minipage}\\
\makebox[3ex]{$^{ S}$}
\begin{minipage}[t]{14cm}
 supported in part by the MINERVA Gesellschaft f\"ur Forschung GmbH, the Israel Science
 Foundation (grant No. 293/02-11.2) and the US-Israel Binational Science Foundation \
\end{minipage}\\
\vspace{30em} \pagebreak[4]

%  references concerning mebers;

\makebox[3ex]{$^{ a}$}
\begin{minipage}[t]{14cm}
also affiliated with University College London,
 United Kingdom\
\end{minipage}\\
\makebox[3ex]{$^{ b}$}
\begin{minipage}[t]{14cm}
now at University of Salerno, Italy\
\end{minipage}\\
\makebox[3ex]{$^{ c}$}
\begin{minipage}[t]{14cm}
now at Queen Mary University of London, United Kingdom\
\end{minipage}\\
\makebox[3ex]{$^{ d}$}
\begin{minipage}[t]{14cm}
also funded by Max Planck Institute for Physics, Munich, Germany\
\end{minipage}\\
\makebox[3ex]{$^{ e}$}
\begin{minipage}[t]{14cm}
also Senior Alexander von Humboldt Research Fellow at Hamburg University,
 Institute of Experimental Physics, Hamburg, Germany\
\end{minipage}\\
\makebox[3ex]{$^{ f}$}
\begin{minipage}[t]{14cm}
also at Cracow University of Technology, Faculty of Physics,
 Mathemathics and Applied Computer Science, Poland\
\end{minipage}\\
\makebox[3ex]{$^{ g}$}
\begin{minipage}[t]{14cm}
supported by the research grant No. 1 P03B 04529 (2005-2008)\
\end{minipage}\\
\makebox[3ex]{$^{ h}$}
\begin{minipage}[t]{14cm}
now at Rockefeller University, New York, NY
 10065, USA\
\end{minipage}\\
\makebox[3ex]{$^{ i}$}
\begin{minipage}[t]{14cm}
now at DESY group FS-CFEL-1\
\end{minipage}\\
\makebox[3ex]{$^{ j}$}
\begin{minipage}[t]{14cm}
now at Institute of High Energy Physics, Beijing,
 China\
\end{minipage}\\
\makebox[3ex]{$^{ k}$}
\begin{minipage}[t]{14cm}
now at DESY group FEB, Hamburg, Germany\
\end{minipage}\\
\makebox[3ex]{$^{ l}$}
\begin{minipage}[t]{14cm}
also at Moscow State University, Russia\
\end{minipage}\\
\makebox[3ex]{$^{ m}$}
\begin{minipage}[t]{14cm}
now at University of Liverpool, United Kingdom\
\end{minipage}\\
\makebox[3ex]{$^{ n}$}
\begin{minipage}[t]{14cm}
now at CERN, Geneva, Switzerland\
\end{minipage}\\
\makebox[3ex]{$^{ o}$}
\begin{minipage}[t]{14cm}
also affiliated with Universtiy College London, UK\
\end{minipage}\\
\makebox[3ex]{$^{ p}$}
\begin{minipage}[t]{14cm}
now at Goldman Sachs, London, UK\
\end{minipage}\\
\makebox[3ex]{$^{ q}$}
\begin{minipage}[t]{14cm}
also at Institute of Theoretical and Experimental Physics, Moscow, Russia\
\end{minipage}\\
\makebox[3ex]{$^{ r}$}
\begin{minipage}[t]{14cm}
also at INP, Cracow, Poland\
\end{minipage}\\
\makebox[3ex]{$^{ s}$}
\begin{minipage}[t]{14cm}
also at FPACS, AGH-UST, Cracow, Poland\
\end{minipage}\\
\makebox[3ex]{$^{ t}$}
\begin{minipage}[t]{14cm}
partially supported by Warsaw University, Poland\
\end{minipage}\\
\makebox[3ex]{$^{ u}$}
\begin{minipage}[t]{14cm}
now at Istituto Nucleare di Fisica Nazionale (INFN), Pisa, Italy\
\end{minipage}\\
\makebox[3ex]{$^{ v}$}
\begin{minipage}[t]{14cm}
now at Haase Energie Technik AG, Neum\"unster, Germany\
\end{minipage}\\
\makebox[3ex]{$^{ w}$}
\begin{minipage}[t]{14cm}
now at Department of Physics, University of Bonn, Germany\
\end{minipage}\\
\makebox[3ex]{$^{ x}$}
\begin{minipage}[t]{14cm}
also affiliated with DESY, Germany\
\end{minipage}\\
\makebox[3ex]{$^{ y}$}
\begin{minipage}[t]{14cm}
also at University of Tokyo, Japan\
\end{minipage}\\
\makebox[3ex]{$^{ z}$}
\begin{minipage}[t]{14cm}
now at Kobe University, Japan\
\end{minipage}\\
\makebox[3ex]{$^{\dagger}$}
\begin{minipage}[t]{14cm}
 deceased \
\end{minipage}\\
\makebox[3ex]{$^{aa}$}
\begin{minipage}[t]{14cm}
supported by DESY, Germany\
\end{minipage}\\
\makebox[3ex]{$^{ab}$}
\begin{minipage}[t]{14cm}
supported by the Bogolyubov Institute for Theoretical Physics of the National
 Academy of Sciences, Ukraine\
\end{minipage}\\
\makebox[3ex]{$^{ac}$}
\begin{minipage}[t]{14cm}
STFC Advanced Fellow\
\end{minipage}\\
\makebox[3ex]{$^{ad}$}
\begin{minipage}[t]{14cm}
nee Korcsak-Gorzo\
\end{minipage}\\
\makebox[3ex]{$^{ae}$}
\begin{minipage}[t]{14cm}
This material was based on work supported by the
 National Science Foundation, while working at the Foundation.\
\end{minipage}\\
\makebox[3ex]{$^{af}$}
\begin{minipage}[t]{14cm}
also at Max Planck Institute for Physics, Munich, Germany, External Scientific Member\
\end{minipage}\\
\makebox[3ex]{$^{ag}$}
\begin{minipage}[t]{14cm}
now at Tokyo Metropolitan University, Japan\
\end{minipage}\\
\makebox[3ex]{$^{ah}$}
\begin{minipage}[t]{14cm}
now at Nihon Institute of Medical Science, Japan\
\end{minipage}\\
\makebox[3ex]{$^{ai}$}
\begin{minipage}[t]{14cm}
now at Osaka University, Osaka, Japan\
\end{minipage}\\
\makebox[3ex]{$^{aj}$}
\begin{minipage}[t]{14cm}
also at \L\'{o}d\'{z} University, Poland\
\end{minipage}\\
\makebox[3ex]{$^{ak}$}
\begin{minipage}[t]{14cm}
member of \L\'{o}d\'{z} University, Poland\
\end{minipage}\\
\makebox[3ex]{$^{al}$}
\begin{minipage}[t]{14cm}
now at Lund University, Lund, Sweden\
\end{minipage}\\
\makebox[3ex]{$^{am}$}
\begin{minipage}[t]{14cm}
also at University of Podlasie, Siedlce, Poland\
\end{minipage}\\

}

%\end{document}

%\end{tabular}

\newpage

%------------------------------------------------------------------------------
%       Text
%------------------------------------------------------------------------------
\pagenumbering{arabic}
\pagestyle{plain}
% ----------------------------------------------------------------------------
%       Introduction
% ----------------------------------------------------------------------------
\section{Introduction}
\label{sec-intro}

The soft hadronic nature of the photon observed in $\gamma p$
collisions~\cite{rmp:50:261,*bauererratum} is well described
by the vector meson dominance model~\cite{anphy:11:1,*prl:22:981},
in which the photon is considered to be a superposition of
vector mesons interacting with the proton. 
Therefore, the energy dependence above the resonance region of the
total $\gamma p$ cross section, \st, is expected to be similar
in form to that of the total hadronic cross sections, $\sigma_{\rm tot}$,
for $pp, \bar{p}p, \pi p$ and $K p$ interactions.

Donnachie and Landshoff~\cite{DL} demonstrated
that the energy dependences of all hadron-hadron total cross sections
may be described by a simple Regge-motivated form,
\begin{equation}
\sigma_{\rm tot} = A \cdot (W^2)^{\alpha_{\pom}(0)-1}
                 + B \cdot (W^2)^{\alpha_{\reg}(0)-1} \, ,
\label{eq-dl}
\end{equation}
where
$W$ is the hadron-hadron center-of-mass energy,
$A$ and $B$ are process-dependent constants,
and
$\alpha_{\pom}(0)$ ($\alpha_{\reg}(0)$) is
process-independent and interpreted as the Pomeron (Reggeon)
trajectory intercept.

This observation together with the interest in estimating the 
total cross sections at high energies, well
beyond the range probed experimentally (for example for $pp$ 
scattering at the LHC or for cosmic-ray physics),
prompted further Regge-type fits of the energy dependence of the total
hadron-proton cross sections~\cite{cudell2,pr:d65:074024}. 
At sufficiently high energies, the power-like behavior 
of the energy dependence is expected to be modified by the 
Froissart bound~\cite{pr:123:1053,*ncim:42:930}  and the total cross section is
expected to behave as $\ln^2(W^2)$. Recent analyses of 
hadron-proton  and photon-proton cross sections indicate 
that already at present energies 
a $\ln^2(W^2)$ dependence is observed~\cite{pr:d70:091901,godbole,mmblock}.
The data from many experiments must be combined in such fits and
the evaluation of the influence of systematic uncertainties is complex.

At the $ep$ collider HERA,
\st can be extracted from $ep$ scattering at very low
squared momentum transferred at the electron vertex,
$Q^2 \lesssim 10^{-3} \gev^2$.
The measurements of the total $\gamma p$ cross section at HERA for 
$W \simeq 200 \gev$~\cite{
ZEUSpaper92,H1paper93,ZEUSpaper94,H1paper95,ZEUSpaper02}
combined with measurements at low $W$ confirmed that the total
photoproduction cross section has a $W$ dependence similar to that of
hadron-hadron reactions.  This similarity extends to virtualities $Q^2$
of the photon up to $\approx 1 \gev^2$~\cite{pl:b487:53}.

This paper presents a determination of the $W$ dependence
of \st from ZEUS data alone, in the range 194--$296\gev$.
This was made possible because in the final months of operation,
the HERA collider was run with constant nominal positron energy,
and switched to two additional proton energies, 
lower than the nominal value of $920\gev$.
Many of the systematic uncertainties arising in the
extraction of \st are now common
and do not affect the relative values of \st at different $W$.
As the Reggeon term is expected to be small,
the function in \myeq{dl} can be simplified to the form
\begin{equation}
  \sigtot =  A' \cdot \left( \frac{W}{W_0} \right)^{\textstyle 2 \epsilon} \, .
 \label{eq-epsfit}
\end{equation}
This is the first extraction of the logarithmic derivative of the
cross section in $W^2$ from a single experiment.

%====================================================================
%        Kinematics
%====================================================================
\section{Kinematics}
\label{sec-kine}

The photon-proton total cross section was measured
in the process 
$e^+p \rightarrow e^+ \gamma p \rightarrow e^+ X$,
where the interacting photon is almost real.
The event kinematics may be described in terms of Lorentz-invariant
variables: the photon virtuality, $Q^2$,
the event inelasticity, $y$, and
the square of the photon-proton center-of-mass energy, $W$,
defined by
\begin{displaymath}
  Q^{2} = -q^{2} = -(k - k^{\prime})^2 \, , \; \; \; \; \; \; \; \;
  y = \frac{p\cdot q}{p\cdot k} \, , \; \; \; \; \; \; \; \;
  W^{2} = ( q + p )^{2} \, ,
  \nonumber
\end{displaymath}
where $k$, $k^{\prime}$ and $p$ are the four-momenta of the
incoming positron, scattered positron and incident proton,
respectively, and $q = k - k^{\prime}$.
These variables can be expressed in terms of the experimentally 
measured quantities 
\begin{displaymath}
  Q^2 = Q^2_{\rm min} + 4 E_e E_e^{\prime} \sin^2\frac{\theta_e}{2}
                                        \, , \; \; \; \; \; \; \; \;
  y = 1 - \frac{E_e^{\prime}}{E_e} \cos^2\frac{\theta_e}{2}  
    \simeq 1 - \frac{E_e^{\prime}}{E_e} \, , \; \; \; \; \; \; \; \;
 W \simeq 2 \sqrt{E_e E_p y} \, ,
  \nonumber
\end{displaymath}
where
\begin{displaymath}
  Q^{2}_{\rm min} = \frac{m^{2}_{e}y^{2}}{1 - y} \, ,
  \nonumber
\end{displaymath}
$E_{e}$, $E_{e}^{\prime}$ and $E_{p}$ are the energies of the 
incoming positron, scattered positron and incident proton, respectively,
$\theta_e$ is the positron scattering angle with respect
to the initial positron direction and $m_e$ is the positron mass.
The scattered positron was detected in a positron tagger close to the beam
line, restricting  $\theta_e$ (and hence $Q^2$) to small values.
The photon virtuality ranged from the kinematic minimum,
$Q^{2}_{\rm min} \simeq 10^{-6}\gev^2$, up to 
$Q^2_{\rm max} \simeq 10^{-3}\gev^2$,
determined by the acceptance of the positron tagger.

The equivalent photon
approximation~\cite{pr:45:729,*zfp:88:612,*jetp:14:1308} relates
the electroproduction cross section to
the photoproduction cross section.
The doubly-differential $ep$ cross section can be written as
\begin{equation*}
\frac{d^{2}\sigma^{ep}(y,Q^{2})}{dy dQ^{2}} =
 \phi(y,Q^{2}) \sigma^{\gamma p}(y,Q^{2}) \; ,
\end{equation*}
where $\phi(y,Q^{2})$ is the doubly differential photon flux.
The longitudinal cross section is small
($\sigma^{\gamma p}_{L} / \sigma^{\gamma p}_{T} < 0.1\%$~\cite{
zfp:c74:297,*Schildknecht:1997,*acpp:b28:2453}), 
and can be neglected.
Then the transverse component of the flux has the form
\begin{equation}
 \phi(y,Q^{2}) = \frac{\alpha}{2\pi}\frac{1}{Q^2} \left(
 \frac{1+(1-y)^{2}}{y} - \frac{2(1-y)}{y} \frac{Q^{2}_{\rm min}}{Q^{2}}
 \right) \, .
\label{eq-ddflux}
\end{equation}
For each of the incident proton energies, $\sigma^{\gamma p}(y,Q^{2})$
has a small variation as a function of $y$ and $Q^2$
over the range of the measurement ($<1.5$\% over $y$ and $<$0.1\% 
over $Q^2$~\cite{
anphy:11:1,*prl:22:981,rmp:50:261,*bauererratum})
and may be taken to be a constant, $\sigtot$.
Thus, the flux may be integrated over the range of measurement to
give a total flux $F_{\gamma}$, which,
when multiplied by the total $\gamma p$ cross section gives $\sigma_{\rm tot}^{ep}$,
the ep cross section integrated over the measured range,
\begin{equation}
\sigma_{\rm tot}^{ep} = F_{\gamma} \cdot \sigtot \; .
\label{eq-eptogp}
\end{equation}

% ----------------------------------------------------------------------------
%       Setup
% ----------------------------------------------------------------------------
\section{Experimental conditions}
\label{sec-setup}

HERA operated with a positron beam energy of
approximately $27.5\gev$ for
all of the data used in this analysis.
The proton beam energies, in chronological order, were
$920\gev$ for the high-energy run (HER),
$460\gev$ for the low-energy run (LER), and
$575\gev$ for the medium-energy run (MER).

\Zdetdesc

\myZtrackingdesc{\myZcoosysfnA}

\Zcaldesc\ 
Timing information from the CAL was available for
identification of out-of-time beam-gas events.
The energy scale of RCAL had an uncertainty of 1\%.

\begin{figure}[p]
\centering
\epsfig{file=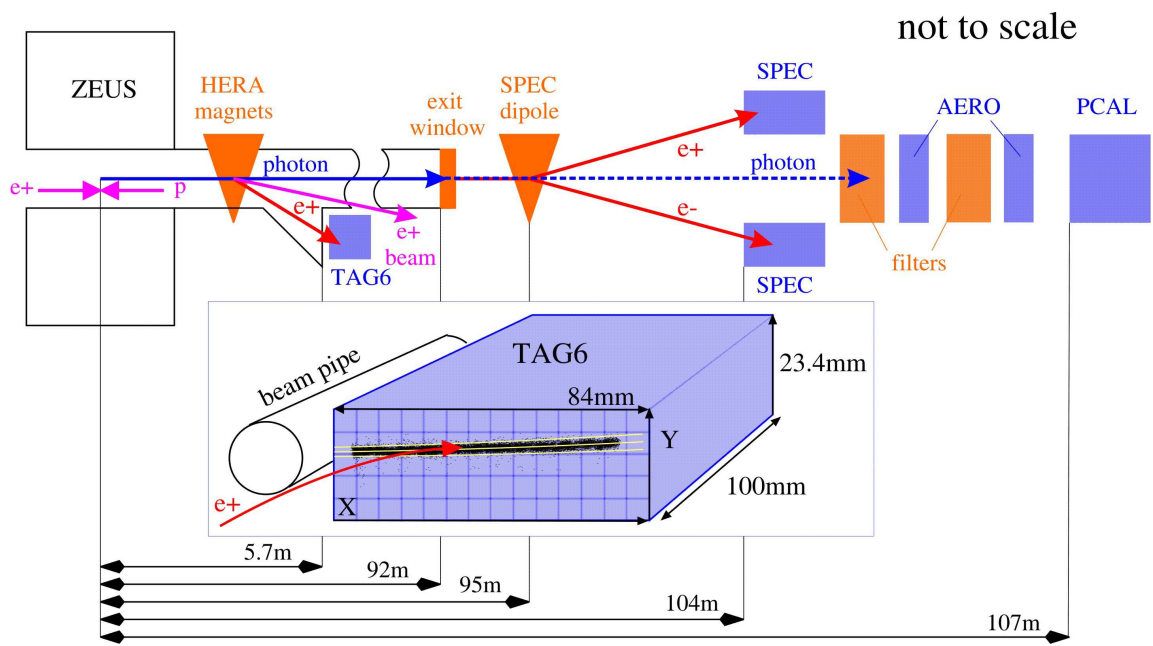,width=15cm}
\caption{
The layout of ZEUS and the luminosity system.
To the right of the TAG6 is a side view, left of
this is a top view.
The inset shows the TAG6 and its cell structure in detail.
Superimposed on the face of the TAG6 is an 
$(X,Y)$ distribution of positrons from a sample of
\BH events from the MER, and the $Y(X)$ selection cuts
described in \Sect{offsel}.
}
\label{fig-layout}
\end{figure}

The luminosity-measuring system consisted of three components.
They were all used for this analysis and are described
in some detail here.
Their layout relative to the ZEUS central detector is
shown in \fig{layout}.

A positron tagger (\ts) was positioned at approximately $Z = -6$ m,
shown in detail in the inset in  \fig{layout}. It consisted of
a tungsten--scintillator spaghetti calorimeter, segmented into
an array of 14 (5) cells with size 6 (4.7) mm in the
horizontal (vertical) direction.
Scattered positrons were bent into it by the first
HERA dipole and quadrupole magnets  after the interaction region,
with full acceptance for positrons with zero transverse momentum 
in the approximate energy range 3.8--$7.1\gev$
with a $y$ range of 0.74--0.86.

At $Z = -92$ m, photons from the interaction point exited the
HERA vacuum system; approximately 9\% of photons converted
into $e^+e^-$ pairs in the exit window.
Converted pairs were separated vertically by a dipole magnet
at $Z = -95$ m.
Pairs from photons in the approximate energy range 15--$25\gev$
were bent into the luminosity spectrometer (SPEC)~\cite{nim:a565:572},
located at $Z = -104$ m.
It consisted of a pair of tungsten--scintillator sandwich calorimeters
located $\approx 10$ cm above and below the plane of
the HERA electron ring.

Photons which did not convert in the exit window were detected
in the lead--scintillator sandwich photon calorimeter
(PCAL)~\cite{acpp:b32:2025}, located at $Z = -107$ m.
It was shielded from primary synchrotron radiation by two carbon
filters, each approximately two radiation lengths deep.
Each filter was followed by an aerogel Cherenkov detector (AERO)
to measure the energy of showers starting in the filters.

The luminosity detectors were calibrated using photons and positrons
from the bremsstrah-lung reaction $ep\rightarrow ep\gamma$.
The SPEC calorimeters were calibrated at the end of HERA fills
by inserting a collimator which constrained the vertical position
of $e^+e^-$ pairs; their energies
were then determined by their vertical positions in the calorimeter
and the magnetic spectrometer geometry.
The energy scale was checked using the endpoint of the \BH
photon spectrum and agreed with the HERA positron beam energy
within 1\%.
The \tss was calibrated using coincidences of \tss positrons
with calibrated SPEC photons and by constraining the sum of
the photon and positron energies to the HERA positron
beam energy~\cite{thesis:schroeder:2008}.
The energy ranges of \BH positrons accepted
by the \tss for different running periods were determined
with uncertainties of 0.01--$0.03\gev$.
The PCAL and PCAL+AERO assembly were calibrated
using coincidences of PCAL(+AERO) photons with calibrated
\tss positrons and constraining the sum of their energies to
the beam energy.

Using photons from the \BH reaction, the luminosity was measured 
independently with the PCAL and with the SPEC.
The systematic uncertainty on the measured luminosity was 1.8\%,
including a relative uncertainty between different running periods of 1\%.
The integrated luminosities used for the \st measurement 
are listed in \tab{table}.

% ----------------------------------------------------------------------------
%       MC
% ----------------------------------------------------------------------------
\section{Monte Carlo simulation}
\label{sec-mc}

Monte Carlo (MC) programs were used to simulate
physics processes in the ZEUS detector.
The {\sc Pythia} 6.416~\cite{pythia2,*pythia} generator was used
for checking the acceptance of the hadronic final state.
The generated events were passed through the
{\sc Geant}~3.21-based~\cite{tech:cern-dd-ee-84-1} ZEUS detector- and
trigger-simulation programs~\cite{zeus:1993:bluebook}. They were
reconstructed and analyzed by the same program chain as the data.
The mixture of photoproduction processes generated by {\sc Pythia}
was adjusted to describe the CAL energy distributions in the
total-cross-section data.
The optimized {\sc Pythia} was also used in the \tss flux
measurement described in \Sect{tagflux}.
That study also used the {\sc Djangoh} 1.6 ~\cite{spi:www:djangoh11}
generator to simulate deep inelastic processes, where the positron
was measured in the CAL.

% ----------------------------------------------------------------------------
%       Selection
% ----------------------------------------------------------------------------
\section{Event  selection}
\label{sec-evsel}

\subsection{Online Selection}

Events for the measurement of \st were collected during
special runs with a dedicated trigger requiring activity
in RCAL and a positron hit in \ts.
The RCAL requirement was a summed energy deposit in the EMC
cells of either more than $464\mev$ (excluding the 8 towers
immediately adjacent to the beampipe) or $1250\mev$ (including
those towers).
The \tss portion of the trigger required at least one cell in
the fiducial region of the tagger to have an energy more than 8 times 
larger than the RMS noise above the pedestal~\cite{thesis:theedt:2009}.
To reduce the background from events with
energy in RCAL and a TAG6 hit caused by a random coincidence with a
\BH event in the same HERA bunch, the energy in the PCAL,
$E_{\rm PCAL}$, was restricted to $E_{\rm PCAL} \lesssim 14\gev$.

\subsection{Offline Selection}
\label{sec-offsel}

Offline, clean positron hits in the \tss were selected by requiring
that the highest-energy cell was not at the edge of the detector.
Showers from inactive material in front of the tagger were rejected
by a cut on the energy sharing among towers surrounding the
tower with highest energy.
The position of the positron was reconstructed
by a neural network trained on an MC simulation
of the \ts~\cite{thesis:gueta:2010}.
The neural-network method was also used to correct the energy
of the positrons for a small number of noisy cells, which were excluded.
Events from the \BH process, selected
by requiring a positron in the \tss in coincidence with
a photon in the SPEC, were used to calibrate the \tss with positrons
with very small transverse momentum.
The energy, $E$, was determined as a function of the horizontal
position, $X$, and the
correlation between $X$ and the vertical position, $Y$, was also measured.
Cuts were placed on $E(X)$ and $Y(X)$ for the photoproduction events to reject
positrons with transverse momentum $p_T \gtrsim 10 \mev$,
off-momentum beam positrons,
and background from beam-gas interactions~\cite{thesis:gueta:2010}.
The $(X,Y)$ distribution of positrons from a sample of \BH events
from the MER,
and the $Y(X)$ cuts, are shown in the inset in \fig{layout}.

In RCAL, the towers immediately horizontally  adjacent  to the
beam-pipe hole had a large rate from off-momentum beam positrons
and debris from beam-gas interactions 
which satisfied the trigger conditions.
In events in which the RCAL cell with highest energy was in one
of these towers, the fraction of total RCAL energy, $E_{\rm RCAL}$, in
that tower was required to be below an $E_{\rm RCAL}$-dependent
threshold~\cite{thesis:gueta:2010}.
This eliminated most of the background and resulted
in only about 2.9\% loss of signal events.

% ----------------------------------------------------------------------------
%       Background
% ----------------------------------------------------------------------------
\section{Data analysis}
\label{sec-bkgsub}

The number of selected events must be corrected to take into account  
beam-gas interactions as well as various effects due to 
random coincidences (overlaps) with \BH interactions. 

Background from positron beam-gas interactions passing the trigger
requirement was determined from non-colliding HERA positron bunches.
This sample was subtracted statistically from the colliding
HERA bunches by the ratio of currents of $ep$ bunches to $e$-only
bunches.
Higher instantaneous luminosity during the HER resulted in
a lower fraction of beam-gas backgrounds relative to the
LER and MER.
The fraction of events subtracted was $\approx 0.2$\% for
the HER and $\approx 1$\% for the LER and MER data samples.

Photoproduction events associated with the \tss hit
could have a random coincidence with
an event in the same HERA bunch from the \BH process,
with the \BH photon depositing more than $14\gev$ in
the PCAL and therefore vetoing the event.
To account for this loss, accepted events were weighted
by a factor determined from the
rate of overlaps at the time the event was accepted.
The fraction of overlaps is proportional to the instantaneous
luminosity, which was higher during the HER relative to the LER and MER.
The correction for this effect was $\approx +2.6$\% for
the HER and  $\approx +1.2$\% for the LER and MER data samples.

Another background came from photoproduction events outside
the $W$ range of the \tss but satisfying the RCAL trigger,
with a random coincidence from \BH hitting the \ts.
The photon from the \BH event may not have been vetoed
by the $E_{\rm PCAL} \lesssim 14\gev$ requirement
due to the limited acceptance and resolution of the PCAL.
Such overlaps were studied using the distribution of the energy
of the PCAL+AERO, $E_{\rm PCKV}$; this offered greatly
improved photon energy resolution over the PCAL alone.
In addition to the \BH events which produced a \tss hit,
this spectrum also contains photoproduction events
associated with the \tss hit overlapping in the same HERA bunch
with a photon from a random \BH event whose positron did not hit the \ts.

\begin{figure}[p]
\centering
\epsfig{file=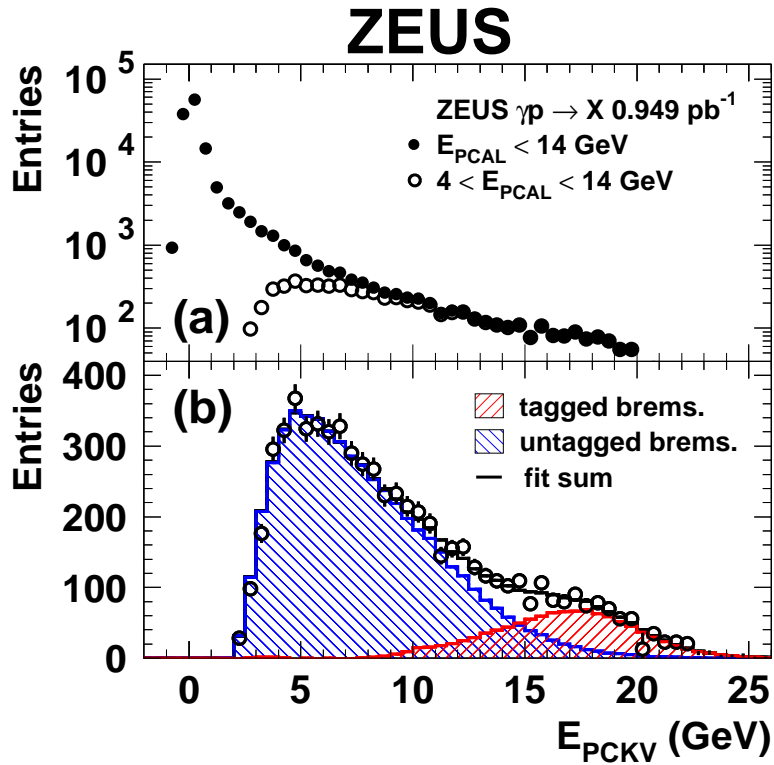,width=10cm}
\caption{
The energy spectrum of photons in the PCAL+AERO;
 (a) the solid points are the MER total-cross-section data
subject to the trigger condition $E_{PCAL}< 14 \gev$; the open points are
subject to the additional condition $E_{PCAL} > 4 \gev$.
 (b) The open points are as above, now shown on
a linear scale. The hatched histograms show the energy spectra of
bremsstrahlung photons with and without a TAG6 requirement. The unshaded
histogram shows the fit of the sum of these two distributions to the
total-cross-section data.
}
\label{fig-epckv}
\end{figure}

The measured $E_{\rm PCKV}$ distribution
from the MER photoproduction data is shown in \fig{epckv}a,
with and without the constraint $E_{\rm PCAL} \gtrsim 4\gev$.
The large peak near $E_{\rm PCKV} = 0$ contains most of the tagged
photoproduction events.
\Fig{epckv}b shows the constrained photoproduction data along
with two distributions from independent samples of 
bremsstrah-lung
events recorded simultaneously with the photoproduction data.
One sample required also the \tss trigger with all 
\tss cuts applied and provides a sample of \tss \BH overlaps.
The other sample was selected with a trigger requiring $E_{\rm PCAL} \gtrsim 4\gev$
and provides a sample of the \BH overlaps independent of a \tss hit.
Only signals from the PCAL were available at the trigger level.
This results in the smeared thresholds in the $E_{\rm PCKV}$
distributions of \fig{epckv}b.
Note that the quoted thresholds in $E_{\rm PCAL}$ are only approximate, since the
trigger conditions were based on uncorrected $E_{\rm PCAL}$ values.
All distributions are restricted to $4\lesssim$$E_{\rm PCAL} \lesssim 14\gev$
to account for the threshold of the various data samples.
The two distributions of \BH events were used to fit the distribution
from the photoproduction events; the component from the tagged
\BH events is the number of tagged \BH overlaps in the sample
where the photon reached the PCAL. The acceptance of photons
in the PCAL was $\approx 85\%$, with losses due to conversions
in the exit window and the limited geometric acceptance from
the aperture defined by the HERA beamline elements.
The number of overlaps seen in the PCAL, corrected for the PCAL acceptance,
is the number of \BH overlaps to subtract from
the selected photoproduction sample.
The uncertainty of 1\% on the PCAL acceptance produces a
systematic uncertainty of $\approx 0.3\%$ on the subtraction,
shown in \tab{table}.

This subtraction procedure was performed in bins of $E_{\rm RCAL}$.
The measured $E_{\rm RCAL}$ distribution before and after the subtraction
 is shown in \fig{ercal}a for the MER sample, together with the
systematic uncertainty from the subtraction procedure.
The amount subtracted is largest at low values of $E_{\rm RCAL}$.
To reduce the statistical and systematic uncertainties
from the subtraction procedure, the signal region for
the \st measurement was restricted to $E_{\rm RCAL}>5\gev$.
The fraction of selected events subtracted was 3.6--4.1\%.
The final numbers of events and their uncertainties
are listed in \tab{table}.

% ----------------------------------------------------------------------------
%       CAL acceptance
% ----------------------------------------------------------------------------
\section{Acceptance of the hadronic final state}
\label{sec-accept}

\begin{figure}[p]
\begin{tabular}{cccc}
 \begin{minipage}{7cm}
 \epsfig{file=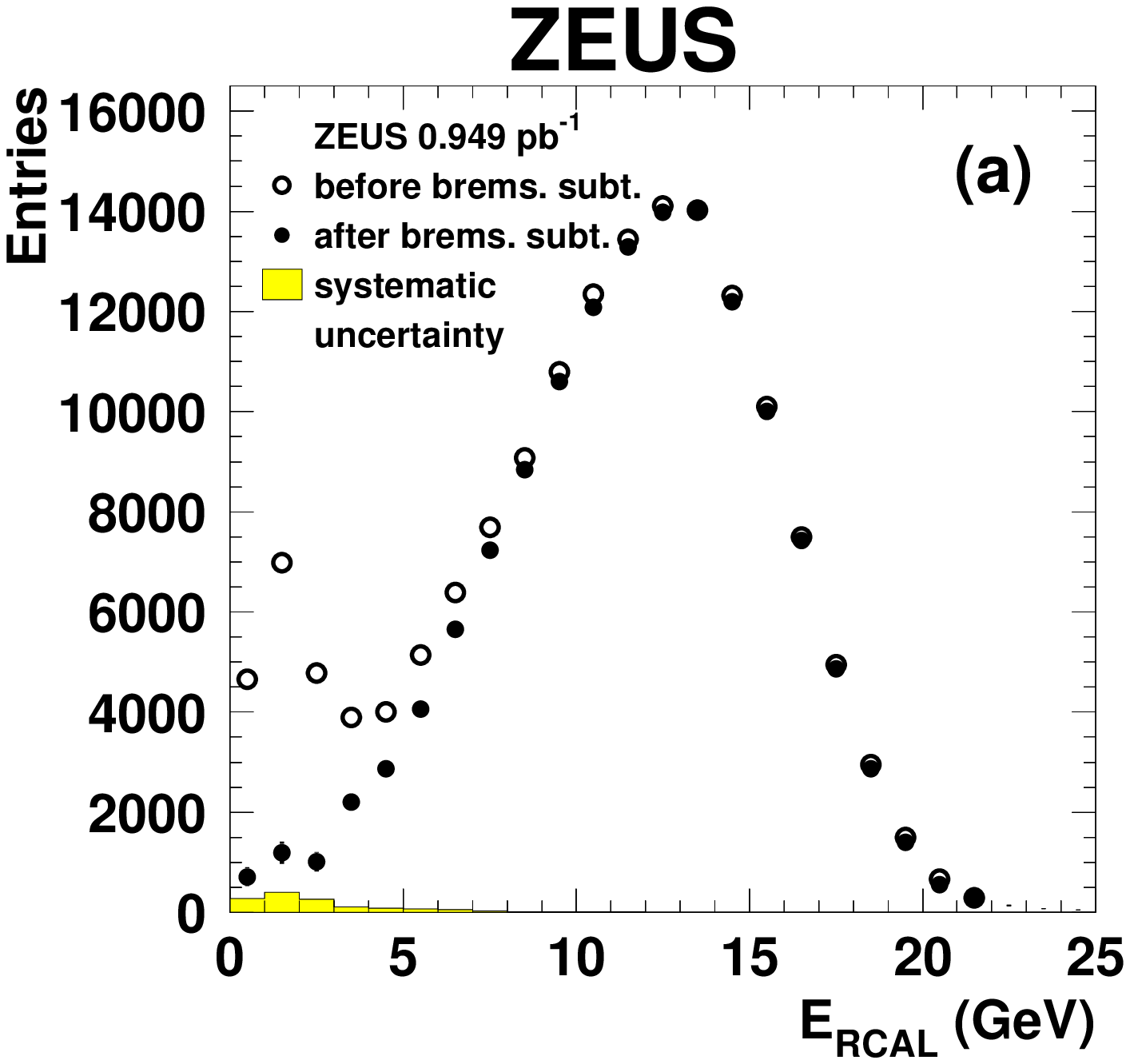,width=7.5cm}
 \end{minipage}
&
 \begin{minipage}{7cm}
 \epsfig{file=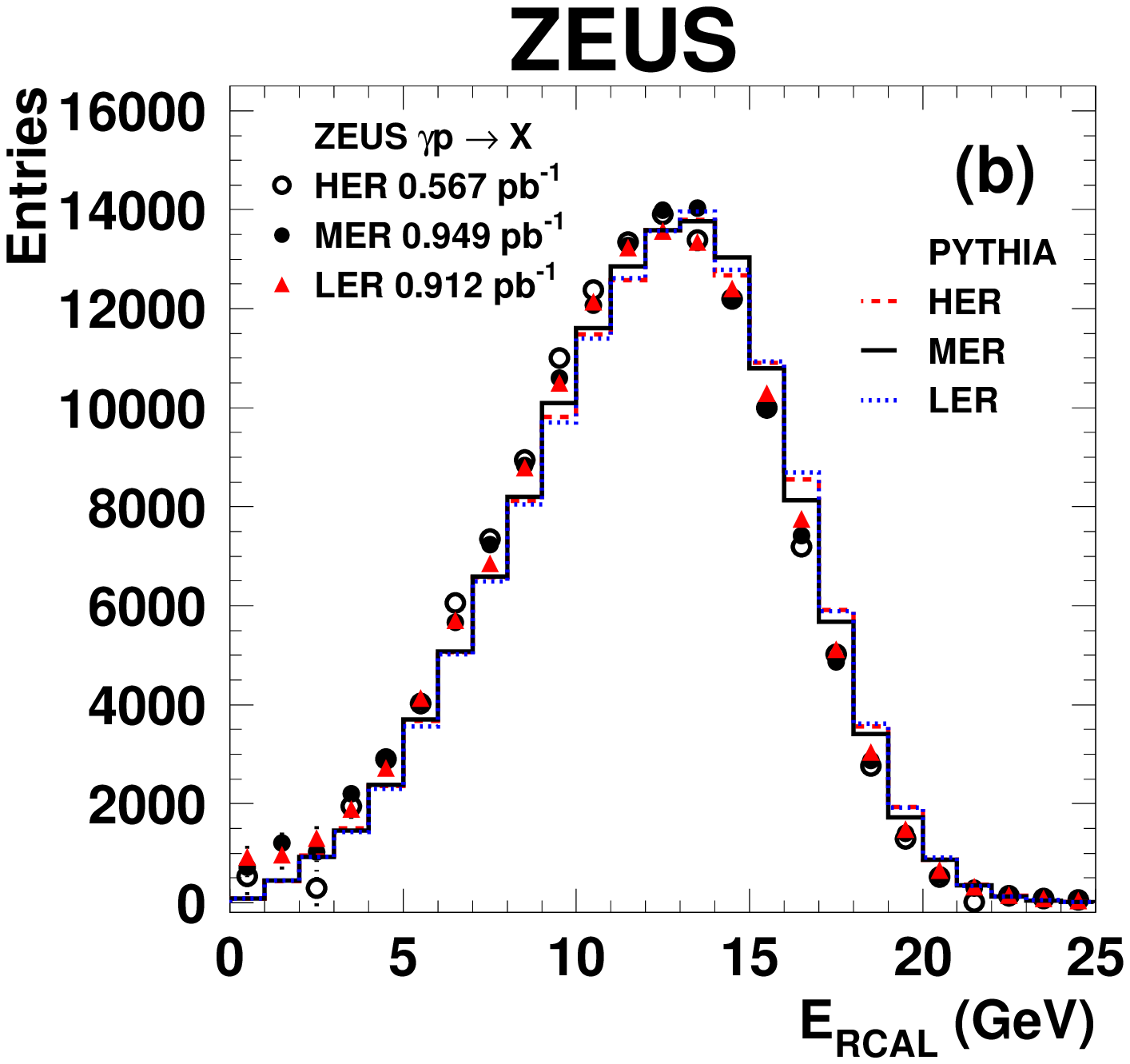,width=7.5cm}
 \end{minipage}
\end{tabular}
\caption{
(a) The $E_{RCAL}$ distribution of the MER sample before and after subtraction
of the TAG6 tagged \BH overlaps.
The shaded histogram shows the systematic uncertainty
of the subtraction procedure, resulting from the
uncertainty on the PCAL acceptance.
(b) The $E_{RCAL}$ distributions after subtraction of \BH overlaps
and the expectations of {\sc Pythia} for all three proton energies.
All distributions are normalized to the MER data for $E_{RCAL} > 5\gev$.
}
\label{fig-ercal}
\end{figure}

The acceptance of the hadronic final state, mainly determined by the
trigger requirement of energy deposit in RCAL, is expected to be the
same for the three energy settings since the positron beam energy, and
thus the photon energy, remained approximately the same throughout.
The trigger
covers the photon-fragmentation region, which is expected to be $W$
independent due to the phenomenon of limiting fragmentation~\cite{pr:188:2159}.
\Fig{ercal}b shows the measured $E_{\rm RCAL}$ distributions, after all
selections and corrections, for all three proton energies. The HER and LER
distributions were normalized to the MER distribution for $E_{\rm RCAL}>5\gev$.
The three distributions are very similar in shape.
The acceptance of the hadronic final state
was further investigated using the {\sc Pythia} MC described in \Sect{mc}.
\Fig{ercal}b shows the $E_{\rm RCAL}$ distribution from the simulation
for all three proton energies, normalized to the MER data for $E_{\rm RCAL}>5\gev$.
The differences between {\sc Pythia} and the data
are similar for all proton energies.
The acceptance for the hadronic final state determined
from {\sc Pythia} was found to be fairly high (above 80\% for most of
the processes) and as expected $W$ independent within small
statistical uncertainties.

% ----------------------------------------------------------------------------
%       t6 flux
% ----------------------------------------------------------------------------
\section{Determination of the photon flux}
\label{sec-tagflux}

The photon flux accepted by the \ts, $\Fts$, is the integral of
the doubly differential flux weighted by the acceptance of the \ts,
$A_{\ts}$, as a function of $(y,Q^2)$
\begin{equation*}
  \Fts = {\textstyle \int} dy dQ^2 \phi(y,Q^2) A_{\ts}(y,Q^2) \, ,
\end{equation*}
where $\phi$ is defined in \myeq{ddflux}.

The HERA magnets closest to the interaction region provided
fields guiding both the proton and positron beams.
Accommodation of the different proton energies required
changes in the fields. These magnets determined the
range of positron energies and scattering angles
accepted by the \ts. 
The changes in accepted kinematic region required a determination 
of the photon flux in \myeq{eptogp} separately for each of
the proton energies.

In order to measure $\Fts$, a sample of photoproduction events
with and without a \tss tag is needed. This was provided by an
independent sample of photoproduction events, selected by a trigger based on
$E-P_Z$, explained in detail below. 
The total $ep$ cross section measured for such a sample is
\begin{equation}
\sigma_{ep}^{\rm tot}= {\textstyle \int} dy dQ^2
 \phi(y,Q^2) \sigma_{\gamma p}(y,Q^2) \Ainc(y,Q^2) \, ,
  \label{eq-sigeptot}
\end{equation}
where $\sigma_{\gamma p}(y,Q^2)$ is the photoproduction cross
section and $\Ainc(y,Q^2)$ is the acceptance for the
selection of the inclusive photoproduction sample.
The $ep$ cross section measured for the subset of this sample
with a \tss tag is
\begin{eqnarray}
\sigma_{ep}^{\ts} &=& {\textstyle \int} dy dQ^2
 \phi(y,Q^2) \sigma_{\gamma p}(y,Q^2) \Ainc(y,Q^2) A_{\ts}(y,Q^2) \nonumber \\
 &=&  \sigma_{\gamma p}^0 \Ainc^0 \Fts \, .
  \label{eq-sigepts}
\end{eqnarray}
The last step follows from the assumption that
$\sigma_{\gamma p}(y,Q^2) = \sigma_{\gamma p}^0$ and
$\Ainc(y,Q^2) = \Ainc^0$
are constant over the small $(y,Q^2)$ region selected by the \ts.
Then, the fraction of selected events with a \tss tag is
\begin{equation}
 r_{\ts} = \frac{\sigma_{\gamma p}^0 \Ainc^0 \Fts}{\sigma_{ep}^{\rm tot}} \, .
\label{eq-rts}
\end{equation}
A MC sample of photoproduction events was then selected
in the same way as these data;
it has the same total $ep$ cross section as in \myeq{sigeptot}.
A well defined test region in $(y,Q^2)$, corresponding to the
\tss region, was used to select a subset of the MC events.
The integrated flux of the test region, $\Ftest$,
was evaluated by integrating the function in \myeq{ddflux} 
numerically over the test region.
The cross section for the events in this region has a
form similar to that of \myeq{sigepts}. The fraction of
selected MC events in the test region is
\begin{equation}
 r_{\rm test} = 
      \frac{\sigma_{\gamma p}^0 \Ainc^0 \Ftest}{\sigma_{ep}^{\rm tot}}
 \, .
\label{eq-rtest}
\end{equation}
Then, from \myeqsand{rts}{rtest}
\begin{equation*}
 \Fts = \frac{r_{\ts}}{r_{\rm test}} \cdot \Ftest \, .
\end{equation*}

The photoproduction data used for this measurement of the \tss flux were
collected simultaneously with the total-cross-section data (LER/MER),
or during a similar running period (HER).
They were collected with a trigger requiring $E-P_Z > 30\gev$,
where $E-P_Z = \sum_i E_i (1-\cos{\theta_i})$,
with the sum running over all CAL cells with
energy $E_i$ and polar angle $\theta_i$.
Offline, $E-P_Z > 31\gev$ was required.
The cut on RCAL towers adjacent to the beam-pipe hole described
in \Sect{evsel} was applied.
A good tracking vertex was required with $|Z_{\rm vtx}| < 25\cm$,
and timing in RCAL, and FCAL if available, was required to be
within $3 \ns$ of that of an $ep$ collision; these cuts
reduced beam-induced backgrounds.
Scattered positrons in events with $Q^2 \gtrsim 1\gev^2$ which hit
the CAL, with $E-P_Z \approx 55\gev$, were identified
using a neural network~\cite{nim:a365:508,*nim:a391:360}; 
events with an identified positron were rejected.

A subsample with a positron in the \tss
was selected following the same procedure described in \Sect{evsel};
the same \BH background correction described 
in \Sect{bkgsub} was applied.
For both the inclusive and tagged samples, a small contribution
from positron beam-gas events was subtracted statistically
in the same manner as described in \Sect{bkgsub}; this amounted
to a 1--2.5\% correction for the inclusive sample.

The {\sc Pythia} and {\sc Djangoh} programs described in Section 4 were
used to produce the MC samples.
The {\sc Pythia} samples were restricted to $Q^2 < 1.5\gev^2$,
and the {\sc Djangoh} samples to $Q^2 > 1.5\gev^2$. The MC events were
selected with the
same criteria as for the data, except for the timing cuts.
The {\sc Pythia} and {\sc Djangoh} samples were added to give the
same fraction of events with and without an identified positron
as in the data.
The \tss test region in {\sc Pythia} had the same $y$ range as the
corresponding data set and $Q^2 < 10^{-3} \gev^2$.

\begin{figure}[p]
\centering
\epsfig{file=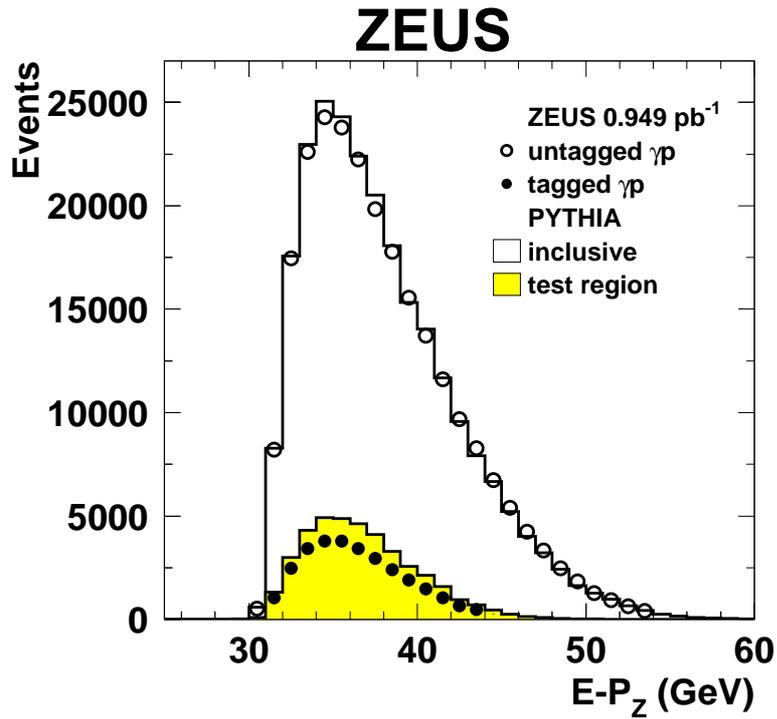,width=10cm}
\caption{
$E-P_Z$ distributions used for the determination of
the photon flux for the MER.
The open points are the photoproduction data collected
with the $E-P_Z>30\gev$ trigger.
The solid points are those data with the additional TAG6
requirement.
The unshaded histogram is the MC simulation with the same
selection, normalized to the photoproduction data for
$35<E-P_Z<50\gev$. The shaded histogram shows
the MC events in the \tss $E_{\gamma}$ range and
with $Q^2 < 10^{-3} \gev^2$.
}
\label{fig-fluxmeas}
\end{figure}

\Fig{fluxmeas} shows the $E-P_Z$ distributions for the MER sample.
Here $E-P_Z$ was calculated using energy-flow 
objects~\cite{epj:c1:81,*thesis:briskin:1998}.
The MC distribution was normalized to the data
in the region $35<E-P_Z<50\gev$.
The MC gives a fair description of the data;
discrepancies between the data and MC are similar for all
three proton energies, and have a negligible
effect on the relative fluxes determined.
The region $35<E-P_Z<50\gev$ was used to determine
the ratios in \myeqsand{rts}{rtest}
for the flux measurement, avoiding trigger-threshold effects
on the low side and unidentified positrons with
$E-P_Z \approx 55\gev$ on the high side.

The experimental data with a \tss tag and the MC in the \tss test region
in \fig{fluxmeas} are for the full $y$ range of the \ts.
The MC shows that there is a change in the acceptance
of inclusive events ($\Ainc(y)$ in \myeq{sigeptot}) across this range,
whereas $\Ainc$ is taken to be constant in \myeq{sigepts}.
To minimize the error of this acceptance variation, the
\tss data were divided into 12 bins according to the horizontal
position of the \tss cell with highest energy; the MC test region
was divided into the corresponding 12 regions of $y$,
based on the \tss $E(X)$ relation described in \Sect{offsel}.
The flux measurement was performed for these 12 regions
and summed. The results are listed in \tab{table}.
The statistical uncertainties on the flux,
dominated by the number of \tss events, are also shown;
the systematic uncertainties are described in the next section.
The flux-weighted mean photon energy was calculated
over the 12 bins. The mean and ranges of photon energies
and $W$ are also listed in \tab{table}.

% ----------------------------------------------------------------------------
%       Sys. Uncert.
% ----------------------------------------------------------------------------
\section{Systematic uncertainties}
\label{sec-sys}

Several sources of systematic uncertainty were investigated besides
the uncertainty on the background subtraction already discussed
in \Sect{bkgsub}.
Any uncertainty correlated for all three proton energies
largely cancels when ratios of cross sections are determined.
The following list provides a summary of the uncertainties and
in parentheses the maximum effects on the ratios of cross sections:

\begin{itemize}

\item
uncorrelated uncertainty on the PCAL acceptance 
affecting the \BH background subtraction: 1\% (0.3\%);

\item
uncertainty on the change of the acceptance of the hadronic final state:
As discussed in \Sect{accept}, the acceptance
has negligible differences for different
center-of-mass energies as it is mostly sensitive to the positron energy
and hence cancels in the ratios of cross sections at different
proton energies.
This variation is ignored here: $<0.1\%$ ($<0.1\%$);
 
\item
uncertainties on the photon flux:

\begin{itemize}
\item uncorrelated statistical uncertainties from
      event samples used for flux determination: 1--1.1\% (1.1\%);
\item uncorrelated uncertainties on the \tss photon energy ranges,
      which result in uncertainties on the flux caused by
      a steep $y$ dependence of $\Ainc(y,Q^2)$
      as discussed in \Sect{tagflux}: 0.01--$0.03\gev$ (1.1\%);
\item correlated uncertainty on the SPEC photon energy scale,
      introducing uncertainties on the flux through
      the $y$ dependence of $\Ainc(y,Q^2)$ in \Sect{tagflux}: 1\% (0.7\%);
\item correlated uncertainty on the CAL energy scale: 1\% (0.5\%);
\item correlated uncertainty on $W$ and $Q^2$ dependences 
      of the photoproduction cross section as modeled in {\sc Pythia},
      determined by varying the power of the $W$ dependence
      and the cutoff mass for $Q^2$~\cite{epj:c7:609}:
                              0.2--2\% ($0.03$\%);
\item uncorrelated uncertainty due to the
      statistical uncertainties in the procedure
      to determine the flux: 1--1.2\% (1.2\%);
\end{itemize}

\item
uncorrelated uncertainty on luminosity as described in 
\Sect{setup}: 1\% (1\%).

\end{itemize}

All uncorrelated systematic uncertainties were added in quadrature;
the largest contributions were from the statistical uncertainties
of the flux determination and the luminosity uncertainty.
The uncertainties are summarized in \tab{table}.

% ----------------------------------------------------------------------------
%       Results
% ----------------------------------------------------------------------------
\section{Energy dependence of the total cross section}
\label{sec-results}

The total photon-proton cross section for one proton energy
is given by
\begin{equation*}
  \sigtot = \frac{N}{\Lumi \cdot \Fts \cdot A_{\rm RCAL}} \, ,
\end{equation*}
where $N$ is the measured number of events, $\Lumi$ is the
integrated luminosity, $\Fts$ is the fraction of the photon flux
tagged by the \ts, and $A_{\rm RCAL}$ is the acceptance
of the hadronic final state for tagged events.

\begin{figure}[p]
\centering
\epsfig{file=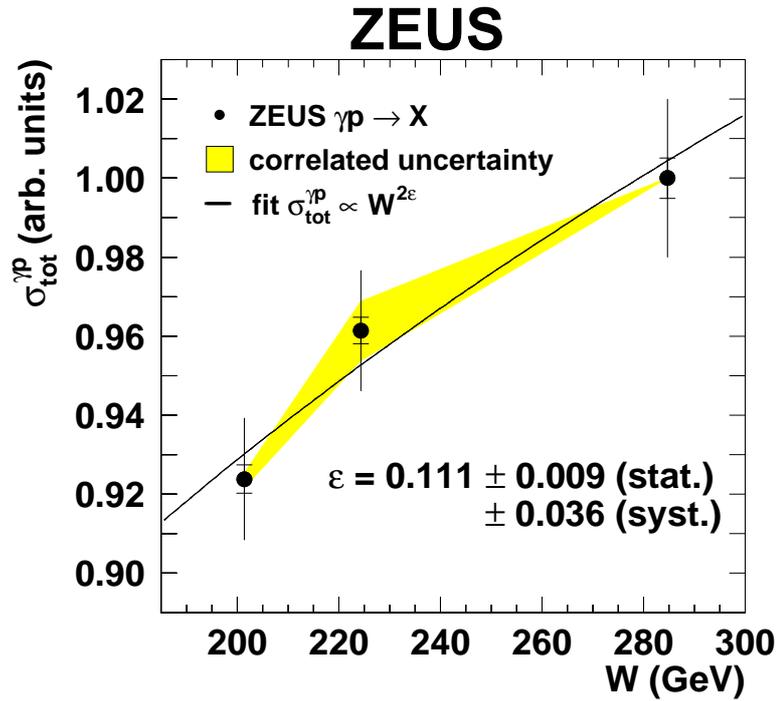,width=10cm}
\caption{
The $W$ dependence of the total photon-proton
cross section, normalized to the value for the HER.
The inner error bars show the statistical uncertainties
of the total-cross-section data; the outer error
bars show those uncertainties and all
uncorrelated systematic uncertainties added in quadrature.
The shaded band shows the effect of the correlated systematic
uncertainties.
The curve shows the fit to the form
$\sigtot \propto W^{2 \epsilon}$.
}
\label{fig-wdep}
\end{figure}

\Fig{wdep} shows the measured relative values of \st as a function of $W$,
where the cross section for HER is normalized to unity.
The functional form of \myeq{epsfit} was fit to the relative cross
sections, with the parameter $W_0$ chosen to minimize correlations
between the fit parameters $A'$ and  $\epsilon$.
The fit was performed using only the statistical uncertainties,
and separately with all the uncorrelated systematic
uncertainties (as in \Sect{sys}) added in quadrature.
The correlated shifts discussed in \Sect{sys} were then applied
to the data and the fit repeated; the change in $\epsilon$
was negligible.
All uncertainties are listed in \tab{table}.
The result for the logarithmic derivative in $W^2$ of 
the energy dependence is
\begin{equation*}
\epsilon = 0.111 \pm 0.009 \, {\rm (stat.)} \pm 0.036 \, {\rm (syst.)} \, .
\end{equation*}

In the picture in which the photoproduction cross section is 
$\propto \ln^2(W^2)$ as required by the Froissart bound~\cite{pr:d70:091901}, 
$\epsilon \approx 0.11$ is expected, in agreement with the present measurement.

The interpretation of this result in terms of the Pomeron intercept is
subject to assumptions on the Reggeon contribution in the
relevant  $W$ range.  If the relative Reggeon contribution,
$B/A$ in \myeq{dl}, is as
assumed in a previous ZEUS analysis~\cite{np:b627:3}, and
$\alpha_{\reg}(0)-1=0.358$~\cite{cudell2},
then $\alpha_{\pom}(0)-1 =\epsilon+0.006$. 
For a relative Reggeon
contribution as measured in another ZEUS analysis~\cite{pl:b487:53}, 
and $\alpha_{\reg}(0)-1=0.5$,
close to the value obtained by Donnachie and Landshoff~\cite{DL},
the Pomeron intercept would be $\alpha_{\pom}(0)-1 =\epsilon+0.002$.

The most recent analysis of all hadronic cross sections
using a fit taking into account Pomeron and Reggeon
terms~\cite{pr:d65:074024}
yielded a Pomeron intercept of 0.0959 $\pm$ 0.0021. This is
in agreement with the result presented here.

% ----------------------------------------------------------------------------
%       Summary
% ----------------------------------------------------------------------------
\section{Summary}
\label{sec-summary}

The energy dependence of the total photon-proton cross section has been
measured using three different center-of-mass energies
in the range $194 \leq W \leq 296\gev$.
A simple $W^{2\epsilon}$ dependence was assumed and a value of
\begin{equation*}
\epsilon = 0.111 \pm 0.009 \, {\rm (stat.)} \pm 0.036 \, {\rm (syst.)}
\end{equation*}
was determined from a fit to the data. This is the first determination
of the energy dependence of the total cross section
at high energy in a single experiment.
The possible Reggeon contribution, though model-dependent,
is expected to be at most a few percent and therefore  the measured
value of $\epsilon$ is compatible with  the energy dependence observed
in hadron-hadron interactions.

\section*{Acknowledgments}
We appreciate the contributions to the construction and maintenance of
the ZEUS detector of many people who are not listed as authors. The
HERA machine group and the DESY computing staff are especially
acknowledged for their success in providing excellent operation of the
collider and the data-analysis environment. We thank the DESY
directorate for their strong support and encouragement.

\vfill\eject

%------------------------------------------------------------------------------
%       Bibliography
%------------------------------------------------------------------------------

\providecommand{\etal}{et al.\xspace}
\providecommand{\coll}{Coll.\xspace}
\catcode`\@=11
\def\@bibitem#1{%
\ifmc@bstsupport
  \mc@iftail{#1}%
    {;\newline\ignorespaces}%
    {\ifmc@first\else.\fi\orig@bibitem{#1}}
  \mc@firstfalse
\else
  \mc@iftail{#1}%
    {\ignorespaces}%
    {\orig@bibitem{#1}}%
\fi}%
\catcode`\@=12

%------------------------------------------------------------------------------
%       Tables
%------------------------------------------------------------------------------

%-------------------------------------------------------------------------------
%       sigtot table
%-------------------------------------------------------------------------------

\begin{table}[p]
\begin{center}
\begin{tabular}{|c|c||c|c|c|}
\hline
 & & LER & MER & HER \\
\hline\hline
 $E_p$ & $\gev$  & 460 & 575 & 920 \\
\hline
 $E_e$ & $\gev$  & 27.50 & 27.52 & 27.61 \\
\hline
 $\Lumi$ & $\nbi$ & 912 & 949 & 567 \\
\hline
 $E_{\gamma}^{\rm min}$ & $\gev$       & 20.49 & 20.29 & 20.42 \\
 $E_{\gamma}^{\rm max}$ &              & 23.66 & 23.60 & 23.81 \\
 $\langle E_{\gamma} \rangle$ &        & 22.04 & 21.88 & 22.03 \\
\hline
 $W^{\rm min}$ & $\gev$       & 194 & 216 & 274 \\
 $W^{\rm max}$ &              & 209 & 233 & 296 \\
 $\langle W \rangle$ &        & 201 & 224 & 285 \\
\hline
 $N$ & events  & 116740 & 128954 & 76310 \\
 $\pm$ stat. & &    457 &    447 &   388 \\ 
 $\pm$ syst. & &    326 &    329 &   224 \\
\hline
 $\Fts$        & $ \times 10^{-3}$ & 0.877 & 0.895 & 0.852 \\
 $\pm$ stat.        &              & 0.009 & 0.009 & 0.010 \\
 $\pm$ uncor. syst. &              & 0.006 & 0.005 & 0.010 \\
\hline
 $\sigtot/\sigtot({\rm HER})$ & &  0.924 &                 0.961 & 1 \\
 $\pm$ stat.        & &            0.004 &                 0.003 & 0.005 \\
 $\pm$ uncor. syst. & &            0.015 &                 0.015 & 0.019 \\
 $\pm$ cor. syst.   & & $^{0.002}_{0.001}$ & $^{0.008}_{0.007} $ & 0 \\
\hline
\end{tabular}
\caption{
Parameters and results for the three proton energies.
For the correlated systematic uncertainties on the relative
cross sections, the LER and MER values shift up and down
by the listed values, while the HER value is fixed to 1.
}
  \label{tab-table}
\end{center}
\end{table}

%------------------------------------------------------------------------------
%       Figures
%------------------------------------------------------------------------------

\clearpage

\clearpage

\clearpage

\clearpage

\clearpage

%
%       ... that's it
%
\end{document}